\documentclass[twocolumn,prb,superscriptaddress,floatfix]{revtex4}

\usepackage{amsfonts}
\usepackage{amssymb}
\usepackage{graphicx}

\begin{document}

\author{Paolo~Michetti$^{1,2}$, Patrik~Recher}
\affiliation{Institute for Theoretical Physics and Astrophysics, University of W\"urzburg, D-97074 W\"urzburg, Germany}

\author{Giuseppe~Iannaccone}
\affiliation{Dipartimento di Ingegneria dell'Informazione: Elettronica, Informatica, Telecomunicazioni, Universit\`a~ di Pisa, Via Caruso 16, 56122 Pisa, Italy}

\email{michetti@physik.uni-wuerzburg.de}

\title{Electric-Field-control of spin rotation in bilayer graphene }

\begin{abstract}
  The manipulation of the electron spin degree of freedom is at the core of the spintronics paradigm, 
  which offers the perspective of reduced power consumption, enabled by the decoupling of
  information processing from net charge transfer. 
  Spintronics also offers the possibility 
  of devising hybrid devices able to perform logic, communication, and storage operations.
  Graphene, with its potentially long spin-coherence length, is a promising 
  material for spin-encoded information transport.
  However, the small spin-orbit interaction is also a limitation for the design of 
  conventional devices based on the canonical Datta-Das spin-FET.
  An alternative solution can be found in magnetic doping of graphene, or, as discussed 
  in the present work, in exploiting the proximity effect between graphene and Ferromagnetic Oxides (FOs).
  Graphene in proximity to FO experiences an exchange proximity interaction (EPI), that acts as an effective Zeeman 
  field for electrons in graphene, inducing a spin precession around the magnetization axis of the FO.
  Here we show that in an appropriately designed double-gate field-effect transistor, with a bilayer
  graphene channel and FO used as a gate dielectric, spin-precession of carriers can be turned ON and OFF with the 
  application of a differential voltage to the gates.
  This feature is directly probed in the spin-resolved conductance of the bilayer.
\end{abstract}

\maketitle

Graphene has attracted much attention since its first experimental fabrication~\cite{novoselov2004}, 
due to its exceptional electronic properties linked to the Dirac physics of its low-energy 
quasiparticles~\cite{castro2009}.
Thanks to its extremely high mobility~\cite{geim2007}, graphene is also a promising material for
nanoelectronics, where however the presence of a semiconducting gap is required. 
One-dimensional graphene-related structures like nanoribbons~\cite{bai2009} 
and carbon nanotubes~\cite{seidel2005} have been employed 
succesfully for nanoelectronic devices, but their practical applications are limited by the need of 
single-atom precision in the definition of their transversal width and radius, respectively.
Carbon-based materials like epitaxial graphene on SiC~\cite{michetti2010} and bilayer 
graphene~\cite{fiori2009} have been shown to be promising for the realization of tunneling 
Field-Effect Transistors (FETs), 
while the gap is not sufficient for conventional FETs~\cite{cheli2009,cheli2010}.

Spin-orbit coupling plays a crucial role in spintronics, providing a way
to manipulate electron spin by means of an external field. 
This is at the heart of most proposed spintronic devices, such as the Datta-Das spin-FET~\cite{datta1990}.
However, theoretical studies have shown that spin-orbit coupling in graphene is extremely small~\cite{min2006, huertas2006}.
Therefore, conventional spintronics mechanisms are not applicable to graphene.
On the other hand, graphene is attractive for spintronics because of its long spin-coherence time~\cite{tombros2007}.
Moreover, simulations show that magnetic doping~\cite{jayasekera2010} in graphene, 
or edge functionalization in GNRs~\cite{son2006,cantele2009}, lead to spin-splitted bands and potentially to a semi-metal 
dispersion relation, particularly attractive for spintronic applications~\cite{zutic2004}.

An alternative approach can come from the exploitation of the interfacial proximity 
with a ferromagnetic oxide.
Indeeed, it has been proposed that a spin splitting (acting as an effective  Zeeman field) can arise in graphene 
due to Exchange Proximity Interaction (EPI) between electrons in graphene and localized electrons in
a FO layer adjacent to the graphene layer~\cite{semenov2007, huegen2008}. 
The effective Zeeman splitting (which has been estimated to be of the order of $5$~meV~\cite{huegen2008}) 
acts on the spin of graphene carriers inducing a precession around the magnetization 
axis of the FO.
The same mechanism was proposed for bilayer graphene, and the modification of the electronic 
structure and of the magnetoresistance as a function of relative angle between 
the magnetization axes of the upper and lower FO spacers has been investigated~\cite{semenov2008}.  
The spin filtering properties of bilayer graphene with multiple \emph{magnetic} barriers with EPI were also investigated in Ref.~\onlinecite{dellanna2009}.
However, the control on the spin-transport properties exerted by an electric field, driven by gates, 
has not yet been the subject of investigation.
\begin{figure}[bh]
  \centering
  \includegraphics[width=7.cm]{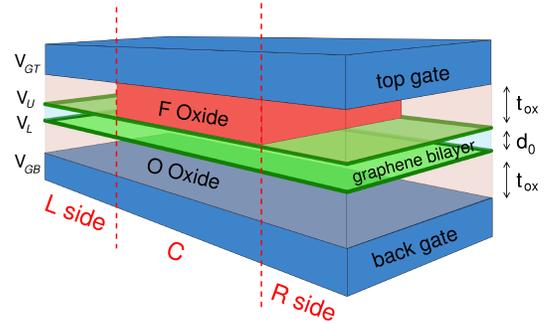}
  \caption{
    (color online) Picture representing a double-gated graphene bilayer structure in which the central part (C) 
    is characterized by the use of a ferromagnetic oxide as spacer between the upper layer and the top gate.
    The oxide thickness $t_{ox}$ and the interlayer distance $d_0$ are indicated.
    The potential of the top gate $V_{GT}$ and the one of the back gate $V_{GB}$ are externally fixed, 
    inducing potential values $V_U$ and $V_{L}$ on the upper and lower graphene layer, respectively, via 
    non-linear Poisson equation.
  }
  \label{fig:0}
\end{figure}

It is possible to induce an energy gap in bilayer graphene by applying an electric field perpendicular 
to the graphene plane~\cite{ohta2006, castro2007, oostinga2008, zhang2009} .
For small kinetic energy, first valence and conduction band wavefunctions are 
driven towards different planes of the bilayer by the applied field.
This means that electrons in the first valence and conduction band have different 
probabilities to be found on the upper and lower plane.
Therefore, we can devise a way to tune the spin-rotation of
carriers in the bilayer graphene by reversing the gate voltage.

We consider a double gate bilayer graphene FET where a FO is used as the insulating layer
between the bilayer graphene channel and the top gate, while an ordinary oxide (OO) is used as insulating 
layer for the bottom gate, as shown in Fig.~\ref{fig:0}.
EPI interaction will mainly affect electrons on the upper graphene layer.
By applying a \emph{direct} or \emph{inverse} differential voltage between the gates, 
we determine whether conduction electrons do or do not feel the EPI interaction, 
and whether the associated wavefunctions are quasi-localized either on the upper or on the lower plane.
Consequently, we are able to switch on or off spin precession.

\section{Model}
We discuss here in very generic terms the difference between an insulating material 
made of an ordinary oxide and of a FO.
In both materials electrons reside in similarly localized  wavefunctions, as it is proper of an insulating material, 
but in a FO they will be also characterized by a majority spin component.
If we place a graphene sheet in proximity to a FO, rather than to a OO, 
in general we can expect a similar contribution for the direct Coulomb interaction between electrons 
in graphene and in the oxide, but a completely different contribution from the exchange interaction.
Indeed, the exchange interaction requires ``exchanged'' electrons to have the same 
spin orientation, and therefore graphene electrons will feel a very different effective 
exchange proximity interaction (EPI) for majority and minority spin components with respect to the FO.
Moreover, while the direct Coulomb interaction is long-ranged, EPI requires an overlap 
of the wavefunctions of ``exchanged'' particles.
For this reason EPI interaction, 
as pointed out in Ref.~\onlinecite{semenov2008}, is essentially limited to the graphene layer 
placed in direct proximity to the FO, and it is negligible on more distant layers.     
\begin{figure}[tbh]
  \centering
  \includegraphics[width=7.8cm]{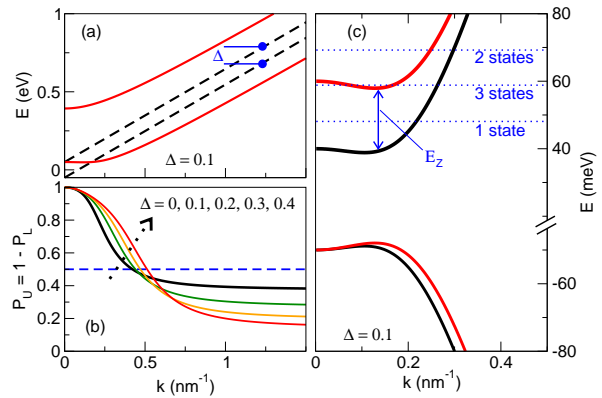}
  \caption{
    (color online) (a) Bilayer dispersion curves (solid lines) compared with the Dirac dispersion curves for the upper 
    and lower graphene planes without inter-plane mixing interaction (dashed line), 
    for $V_U=0.05$~eV and $V_L=-0.05$~eV ($\Delta = 0.1$~eV).
    (b) Projection of the conduction band states onto the upper plane as a function of $k$, 
    for increasing $\Delta$. 
    The black solid line corresponds to the case in (a).
    (c) Dispersion curve for a bilayer system with EPI interaction on the upper plane with $E_{Z}=20$~meV. 
  }
  \label{fig:1}
\end{figure}

We assume here the simplest situation, in which a thin FO layer is deposited between the upper 
graphene plane and the top gate, with magnetization $\vec{M}$, and an OO layer is instead used as insulator 
between the lower graphene layer and the back gate (Fig.~\ref{fig:0}).
The electronic states of bilayer graphene can be described, near the
$K$ point, by the following Hamiltonian~\cite{mccann2006}
\begin{equation}
  H\hspace{-0.08cm}=\hspace{-0.08cm} H_0\hspace{-0.05cm} +\hspace{-0.05cm} H_m\hspace{-0.08cm} =\hspace{-0.05cm} V_0 \mathbb{I}\hspace{-0.05cm} + \hspace{-0.05cm}\left(\begin{array}{llll}
    \frac{\Delta}{2}\hspace{-0.05cm} +\hspace{-0.05cm}h_m & \hat{\pi} & t_\perp & 0 \\
  \hat{\pi}^\dag & \frac{\Delta}{2}\hspace{-0.05cm}+\hspace{-0.05cm}h_m & 0 & 0 \\
  t_\perp & 0 & - \frac{\Delta}{2} & \hat{\pi}^\dag \\
  0 & 0 & \hat{\pi} & - \frac{\Delta}{2}
  \end{array}\right),
\label{eq:hamiltonian}
\end{equation}
where $ V_0=-q(V_U+V_L)/2$ and $\Delta=-q(V_U-V_L)$, with $V_U$ and $V_L$ the upper and lower layer potential, 
respectively, and $q$ is the value of the absolute elementary charge.
In the future, let us denote with the index $U$ the variables relating to the upper (U) layer, and with the index $L$ those relating to the lower (L) layer.
$\hat{\pi} = v_F (p_x +i p_y)$ is the kinetic energy operator (with $p_y\rightarrow - p_y$ for the $K'$ valley), 
$H_m$ is an effective energy term due to the EPI with the ferromagnetic insulators.
We use the parameters $t_\perp=0.39$~eV~\cite{nilson2008, li2009}, i.e. the bilayer interplane coupling, and $v_f\approx 10^{6}$~m/s~\cite{barbier2009}.
Other interlayer coupling terms are neglected, 
in the spirit of Refs.~\onlinecite{huegen2008, semenov2008, barbier2009}, as they would not change the qualitative 
features of the phenomenon described in this work.
The Hamiltonian acts on wavefunctions of the form
\begin{equation}
  \Psi = \left(
  \begin{array}{l}
    \chi_A \\ \chi_B \\ \chi_{B'} \\ \chi_{A'}
  \end{array}
 \right) \frac{e^{i k_x x} e^{i k_y y}}{\sqrt{L_x L_y}},
\label{eq:wavefunction}
\end{equation}
where $A$, $B$ refer to the two inequivalent carbon atoms on the upper
graphene layer, $A'$,$B'$ to that of the lower layer.
$L_x$ and $L_y$ are the channel dimensions along $X$ and $Y$ directions.
Now we distinguish the two spin components along the $Z$ axis,
perpendicular to the plane, therefore $\chi_X$, with
$X=A$, $B$, $A'$, $B'$, has to be regarded as a two-component spinor
\begin{equation}
  \chi_X = \left(
  \begin{array}{l}
    \phi_{X\uparrow} \\ \phi_{X\downarrow}
  \end{array}
 \right).
\end{equation}
The elements $V_U$,$V_L$, $\pi$, $t_\perp$ are diagonal in the spinor
space, while off-diagonal terms can be due to the presence of an
effective Zeeman field.
If we imagine to put the upper graphene layer in contact with a
ferromagnetic insulator having a polarization on the $XY$ plane, 
the exchange interaction gives rise to an off-diagonal coupling 
in the spinor space of the kind
\begin{equation}
  h_m = \alpha \vec{M} \vec{S} = \frac{E_{Z}}{2}\left(\begin{array}{ll}
    0 & m_x - i m_y \\
    m_x + i m_y & 0 
  \end{array}\right),
\end{equation}
where $\hat{M}=(m_x,m_y)$ is the versor of the magnetization vector $\vec{M}$.
For simplicity, we assume for the upper layer a similar effective
Zeeman coupling $h_m$ for the $A$ and $B$ sites, while EPI
vanishes on the lower plane sites $A'$ and $B'$.

\section{Bilayer Graphene}
The idea of controlling spin rotation of bilayer graphene is essentially based on 
the plane-localization properties of the bilayer spinors, when a vertical field is applied.
In particular, let us assume for a moment no EPI interaction, i.e. $\alpha=0$ ($H = H_0$).
In this case the eigenvalues near the $K$, or $K'$, point, are given by the formula~\cite{mccann2006}
\begin{eqnarray}
  (E_{\eta})^{\pm}\hspace{-0.05cm} =\hspace{-0.05cm} V_0\hspace{-0.05cm} \pm\hspace{-0.05cm} \sqrt{\varepsilon_k^2
  +\hspace{-0.05cm} \eta t_\perp \sqrt{\frac{t_\perp^2}{4}\hspace{-0.05cm} + \hspace{-0.05cm}\left(\frac{\Delta^2}{t_\perp^2}\hspace{-0.05cm}+\hspace{-0.05cm}1 \right)\hspace{-0.05cm} \hbar^2 v_f^2 k^2}},
\label{eq:disp}
\end{eqnarray}
with $\varepsilon_k^2 = \frac{\Delta^2}{4} + \hbar^2 v_f^2 k^2+ \frac{t_\perp^2}{2}$, and with $\eta=\pm 1$ for the first ($\eta=-1$) and second ($\eta=+1$) conduction ($+$) or valence($-$) band.  
The bilayer spinors, for a given $\vec{k}$ and energy $E$, are obtained by solving the linear system  $(H_0-E)\Psi=0$.
In Fig.~\ref{fig:1}(a) we show the bilayer dispersion curve, compared with the graphene 
dispersion curves $E_{+}^{\pm} = \pm \hbar v_f k + \frac{\Delta}{2}$ 
and $E_{-}^{\pm} = \pm \hbar v_f k - \frac{\Delta}{2}$, obtained by Eq.~\ref{eq:disp} by decoupling the two layers ($t_\perp=0$).

In Fig.~\ref{fig:1}(b) we plot the projection of the first bilayer conduction band states 
($P_U = |\chi_A|^2 + |\chi_B|^2$) on the U plane. 
The behavior of the projection on the U plane can be easily understood from the bilayer dispersion curve.
In fact, at $k=0$, states of the first bilayer conduction band stand on the Dirac point of the U graphene layer.
At $k=0$, the $A$ and $B$ sublattices are not coupled by the kinetic term $\hat{\pi}$ in the 
Hamiltonian in Eq.~\ref{eq:hamiltonian}, therefore $B$ and $A'$ sublattices are 
not mixed by $t_\perp$ and retain their original character.
Correspondingly, the spinor of the first conduction band will have a $100\%$ weight on the $B$ sublattice.
With increasing $k$, the bilayer spinors have mixed contributions from the two planes 
and eventually, at sufficiently large $k$, $P_U$ tends to a constant value.
An explanation for this comes from the fact that for large $k$, the bilayer conduction bands 
essentially originate from the mixing of only the n-type part of the Dirac cones
 for the U and L graphene sheets (shown in Fig.~\ref{fig:1}(a)), while contributions from the p-like part may be neglected.
This leads to a two-level system with a fixed energy separation of $\Delta$, 
as plotted in Fig.~\ref{fig:1}(a), and fixed tunnel-coupling $\frac{t_\perp}{2}$.     
The solutions of the two level system are $E_\pm = \hbar v_f k \pm \sqrt{\frac{\Delta^2}{4} +\frac{t_\perp^2}{4}}$, 
which correspond to the asymptotic behavior of bilayer conduction bands for large $k$, and 
\begin{equation}
  P_U=\frac{1}{1+\left(\frac{\Delta}{t_\perp}-\sqrt{1+\frac{\Delta^2}{t_\perp^2}}\right)^2}
\end{equation}
for the first conduction band ($\eta=-1$), which explains the plateau in Fig.~\ref{fig:1}(b) at large $k$.
We note that, for a given $k$, $P_U$ of the first conduction band corresponds to $P_L$ of the first 
valence band, and by reversing the potential of both layers one would perfectly exchange 
the projection properties of the two bands.

In Fig.~\ref{fig:1}(c), we plot the first conduction and valence band of a bilayer graphene subjected 
to EPI interaction as described by the Hamiltonian Eq.~\ref{eq:hamiltonian}, with $E_{Z}=20$~meV.
When the EPI interaction is taken into account, electronic wavefuctions traveling on the U plane 
are subject to an effective Zeeman interaction, basically proportional to $P_U$, 
that results in a spin splitting of the bilayer bands by $P_U E_{Z}$.
This proportionality clearly appears in Fig.~\ref{fig:1}(c), where we plot the conduction and valence bands
of a bilayer graphene with a FO deposited on top of the U layer, 
considering a potential energy difference of $\Delta=0.1$~eV between graphene layers.
If we reverse $\Delta$, which can be realized by inverting the bias of top and back gates, 
the spin splitting of conduction and valence bands is inverted, as well as $P_U \leftrightarrow P_L $, 
and the Zeeman splitting at small $k$ will vanish in the conduction band.
In a regime in which small $k$ states are responsible for transport through a FO-contacted 
bilayer region, we will therefore have a degree of control over the electron spin rotation 
induced by the effective Zeeman field.
\begin{figure}[tbh]
  \centering
  \includegraphics[width=7.5cm]{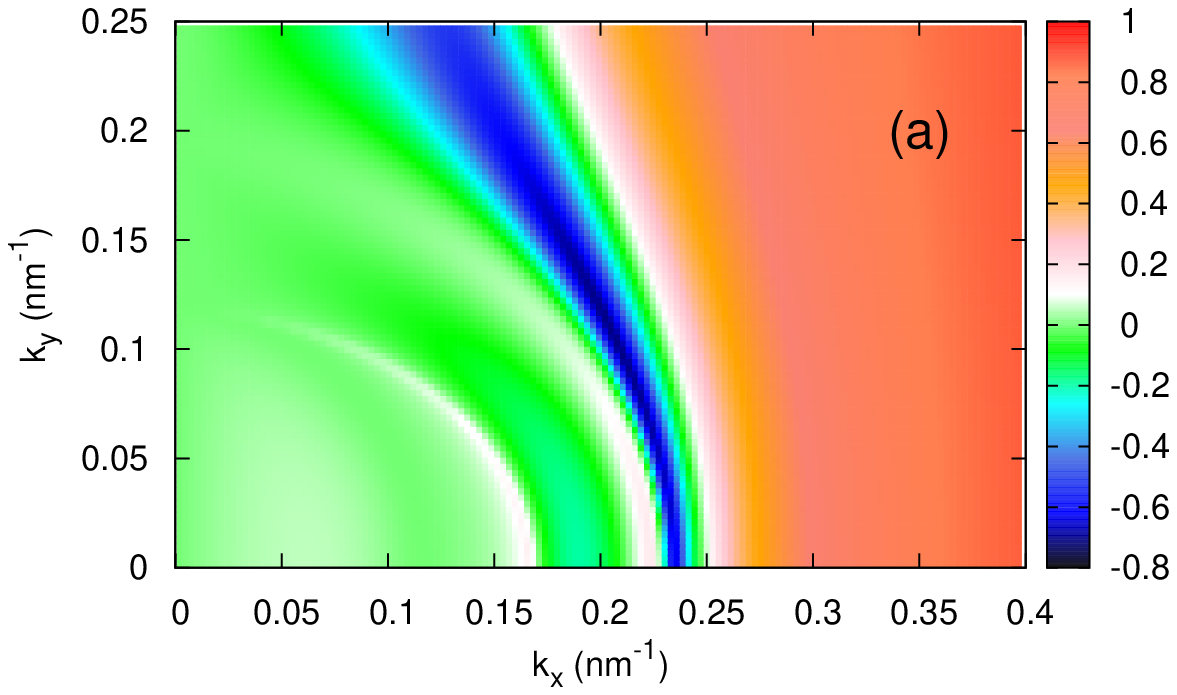}
  \includegraphics[width=7.5cm]{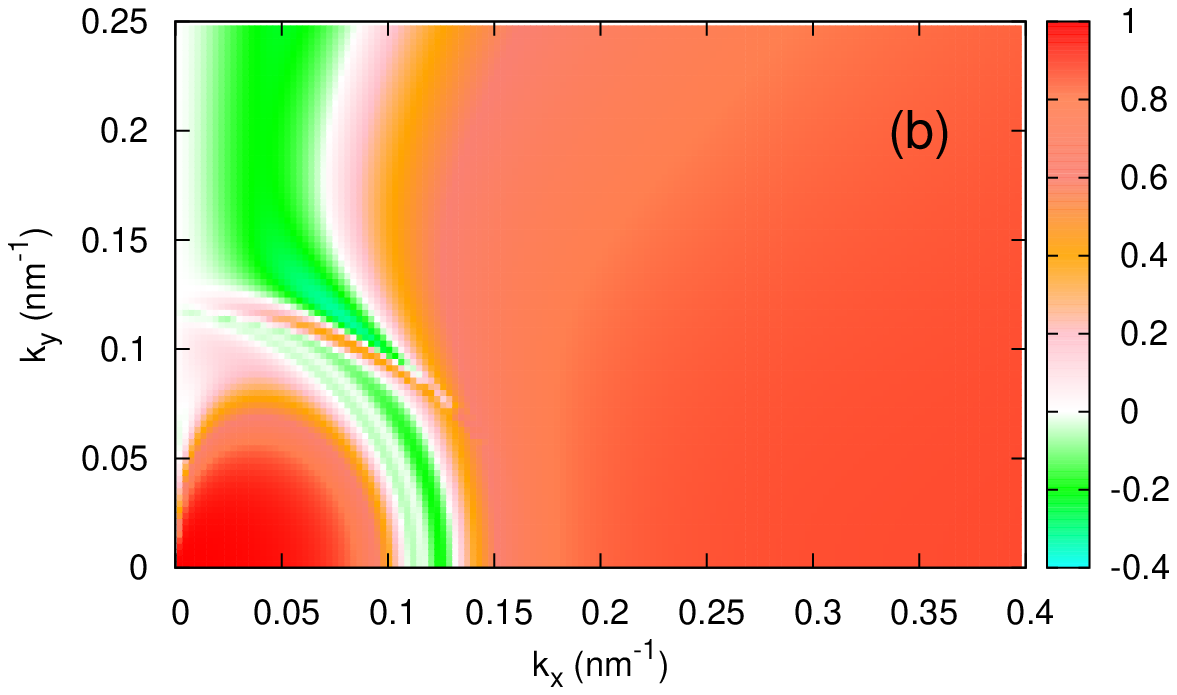}
  \caption{
    (color online) (a) Spin differential transmission $T_s=T_{\uparrow\uparrow}-T_{\uparrow\downarrow}$ through the C region 
    of $L_C= 150 a_{C-C}$ with EPI and $\Delta=0.1$~eV, as a function of the wavevector of the incoming particle.
    (b) $T_s$ for a \emph{reverse} potential energy difference of $\Delta=-0.1$~eV.
  }
  \label{fig:tunn}
\end{figure}

\section{Transmission through a FO gated region}
To calculate the spin rotation properties of our system we analyze the 
transmission through a region of the bilayer in which the EPI coupling is active, as is the case in Fig.~\ref{fig:0}.
For simplicity, we imagine abrupt boundary conditions such that the contact
with the FO is limited to the upper plane of the graphene bilayer from $x=0$ to $x=L_c$.
We consider an incoming conduction band electron from the left side (LS) ($x<0$) 
of given $\vec{k}$ and therefore energy $E_k$, with a chosen $\uparrow$ spin polarization.
Elastic transmission through the active EPI zone conserves $k_y$, because of the space homogeneity 
along the $Y$ axis, but not the spin, and leads to reflected and transmitted components to the left  
and right side (RS) respectively (of $\uparrow$ and $\downarrow$ spin character).

In particular, the LS and RS are described by the Hamiltonian Eq.~\ref{eq:hamiltonian}, 
with $E_{Z}=0$.
We can here find, disregarding the spin which is here conserved, four possible values of the wavevector $k_x$ 
compatible with $k_y$ and energy $E$: $k_x$ and $-k_x$, which are propagating modes, 
$\tilde{k_x}$ and $-\tilde{k_x}$, which can correspond alternatively to propagating modes or to 
evanescent modes, with a finite imaginary part~\cite{barbier2009}.
The number of propagating modes (with real wavevector) corresponds to the number of intersection points 
of the conduction bands with the horizontal line $E=E_k$ on Fig.~\ref{fig:1}(a). 
The remaining modes are evanescent.
Therefore the total wavefunctions on the LS and RS can be written as  
\begin{eqnarray}
  \Psi_L(0) &=& \mathbb{C}_L \vec{r}+ \mathbb{C}_{IN}\vec{s},\\ 
  \Psi_C(L) &=& \mathbb{C}_R \vec{t}  
  \label{eq:vector}
\end{eqnarray}
where the matrices $\mathbb{C}_L$, $\mathbb{C}_R$ and $\mathbb{C}_{IN}$ are built from the spinor set 
of the bilayer system in Eq.~\ref{eq:hamiltonian} without EPI.
$\vec{s}$ is the spin-polarization of the incoming particle and is a 
vector describing the component up and down with respect to the Z axis.
Our calculation starts from fully polarized incoming particles, for which $\vec{s}=(1,0)^T$.   
For all matrices, rows run over the sublattices 
$\{i=A\uparrow,A\downarrow,B\uparrow,B\downarrow,A'\uparrow,A'\downarrow,B'\uparrow,B'\downarrow\}$, 
while columns run over the left region output modes 
$\{ j= -k_x\uparrow, -k_x\downarrow,-\tilde{k_x}\uparrow,-\tilde{k_x}\downarrow \}$ for $\mathbb{C}_L$, 
right region output modes $\{j= k_x\uparrow, k_x\downarrow, \tilde{k_x}\uparrow, \tilde{k_x}\downarrow\}$ 
for $\mathbb{C}_R$, and incoming modes $\{j= k_x\uparrow, k_x\downarrow\}$ for $\mathbb{C}_{IN}$.
The output coefficients are collected in
\begin{eqnarray}
  \vec{r} &=& (r_1,r_{-1}, \tilde{r}_1, \tilde{r}_{-1})^T\\
  \vec{t} &=& (t_1,t_{-1}, \tilde{t}_1, \tilde{t}_{-1})^T 
\end{eqnarray}
with the tunneling coefficients $t_{1}, t_{-1}, \tilde{t}_{1}, \tilde{t}_{-1}$ 
for allowed modes in up or down spin orientation, and the reflection coefficients defined in a 
similar manner as $r_{1}, r_{-1}, \tilde{r}_{1}, \tilde{r}_{-1}$.

In the central part of the system we have a mixing of spin components induced by the effective Zeeman splitting.
Solving the secular equation for the Hamiltonian in Eq.~\ref{eq:hamiltonian} for a given energy $E$ 
and in-plane momentum $k_y$, we obtain $8$ solutions for the wavevector $k_x=\alpha_n$ with $n=1$, $2$ \dots $8$.
The corresponding modes are described by the spinor $\Psi_{\alpha_n}^M$ where the $M$ index is used 
to specify that these states are for the system with Zeeman interaction.
The scattering state in the central part of the system can be generally expressed as
\begin{equation}
  \Psi_C = \sum_n a_n \Psi_{\alpha_n}^M = \mathbb{C}_C \vec{a}. 
\label{eq:c}
\end{equation}

The Dirac equation requires the continuity of spinors at the boundary $x=0$ and $x=L$, which is now 
expressed by the following linear relations
\begin{eqnarray}
  \mathbb{C}_C \vec{a} = \mathbb{C}_L \vec{r} + \mathbb{C}_{IN}\vec{s},\\ 
   \mathbb{C}_R \vec{t} = \mathbb{C}_C \mathbb{P} \vec{a},
  \label{eq:conditions}
\end{eqnarray}
with $\mathbb{P}$ describing the phase accumulation of the different components of the 
scattering state by traveling through the C region.
After elimination of $\vec{a}$, the problem is reduced to the solution of a linear system 
of the kind $\mathbb{M} \vec{x}=\vec{S}$, with 
$\vec{x} = \left(\begin{array}{l} \vec{t}\\ \vec{r} \end{array}\right)$, which can be easily 
solved by standard numerical techniques.
In practice, a source term $\vec{S}$, describing the incoming particle, 
pumps the linear system described through the dynamical matrix $\mathbb{M}$, 
which carries all the information about the transmission through the central region,
and determines the output steady state described by $\vec{x}$.

We consider the transmission of our system, which is given by the sum of the 
outgoing propagating components in the RS.
We choose FO with magnetization along $Y$, so that in the C region we have spin-splitted bands (Y-SSB), 
eigenstates of $S_y$, while we inject and detect in the LS and RS electron spin-polarized along $Z$.
Note that as explained before, with \emph{direct} gate bias, the conduction bands will be spin-splitted, 
while for a \emph{reverse} bias the spin splitting is negligible. 
In Fig.~\ref{fig:tunn}, we show the spin differential transmission 
$T_s=T_{\uparrow\uparrow}-T_{\uparrow\downarrow}$ through the central region with FO deposited on 
the U layer, with $E_{Z}=20$~meV and $L_C=150 \, a_{C-C}$.
$T_s$ is calculated as a function of the wavevector of the incoming particle in the LS.
Fig.~\ref{fig:tunn}(a) is calculated with a \emph{direct} potential energy difference between the graphene planes of 
$\Delta=0.1$~eV.
Indeed a marked resonance is observed with negative values of $T_s$. 
Electrons of this wavevector are transmitted through the barrier with a spin rotation of $\pi$.
In Fig.~\ref{fig:tunn}(b), where $\Delta=-0.1$~eV is \emph{reversed}, such a feature is absent 
and electrons preferentially preserve their spin orientation. 
The spin-transmission properties are therefore dramatically affected by changing between \emph{direct} 
and \emph{reverse} bias of the T and B gates.    
In particular this resonance falls into the mexican hat region of the upper Y-SSB (Fig.~\ref{fig:1}(c)), 
where three propagating states are active: two from the upper Y-SSB and one from the lower one.
The resonance condition is given by the existence of two propagating modes, 
one from each of the two Y-SSBs, 
for which $\Delta k_x L_c = \pi + 2n\pi$, with $n=0,1,..$ (here $n=0$ applies).
In fact, an incoming particle, spin-polarized along $Z$, can be transmitted in the C region 
as a linear combination of two states from the two bands (eigenstates of $S_y$).
These components, traveling through the C region, acquire a net phase difference of $\pi$, which corresponds 
to a net spin-flip process.
Note that the mexican hat-like dispersion makes it possible to have two propagating states with large $\Delta k_x$, 
allowing the fulfillment of the resonance condition with $L_C$ as small as $20$~nm.
Of course, choosing different values for $L_c$ leads to different positions of the spin-flip transmission resonance.

For an incoming particle of lower energy, only the lower Y-SSB contributes  
propagating components in the C region as shown in Fig.~\ref{fig:1}(c).
The overall transmission probability $T=T_{\uparrow\uparrow}+T_{\uparrow\downarrow}$ has an upper limit of $0.5$, 
because the propagating component is eigenstate of $S_y$, and can be seen as a combination of half and 
half $S_z$ spin components.
For the same reason,  the spin differential transmission is close to zero.
For an incoming particle of energy above the resonance, instead, there is one propagating component for each of the 
Y-SSBs (see Fig.~\ref{fig:1}(c)).
However $\Delta k_x$ between these components is much smaller with respect to the resonance 
case and their phase difference accumulated by traveling through the $C$ region is negligible.
Therefore both the spin differential transmission and the overall transmission are close to unity.
\begin{figure}[tbh]
  \centering
  \includegraphics[width=7.5cm]{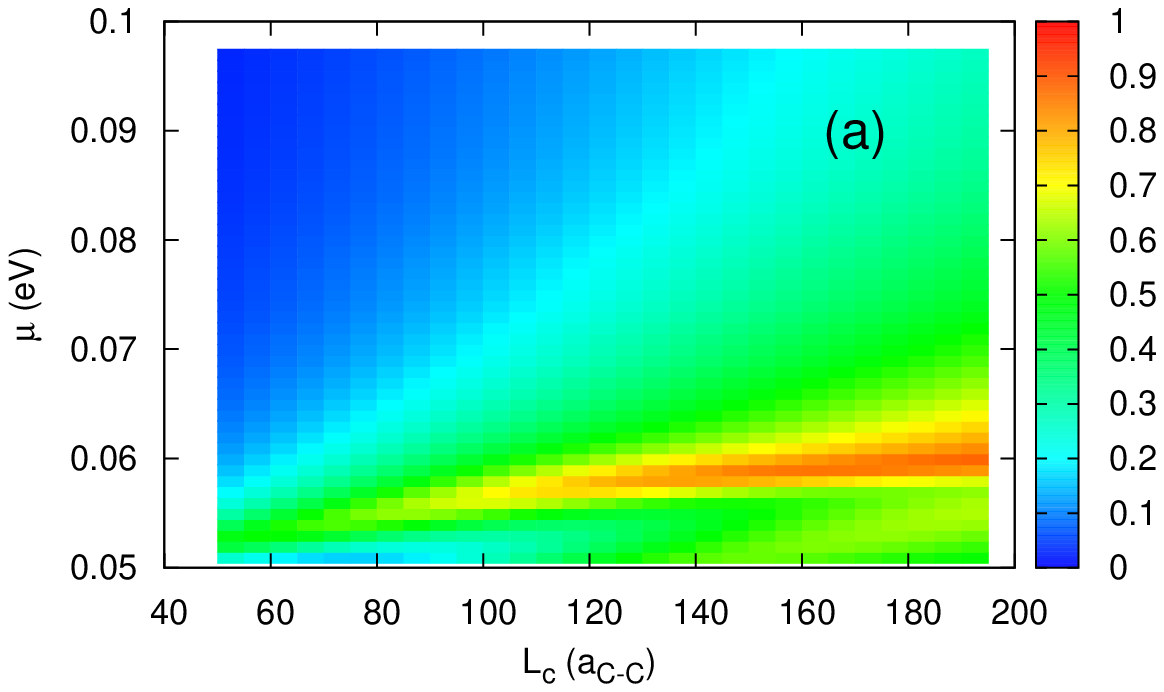}
  \includegraphics[width=7.5cm]{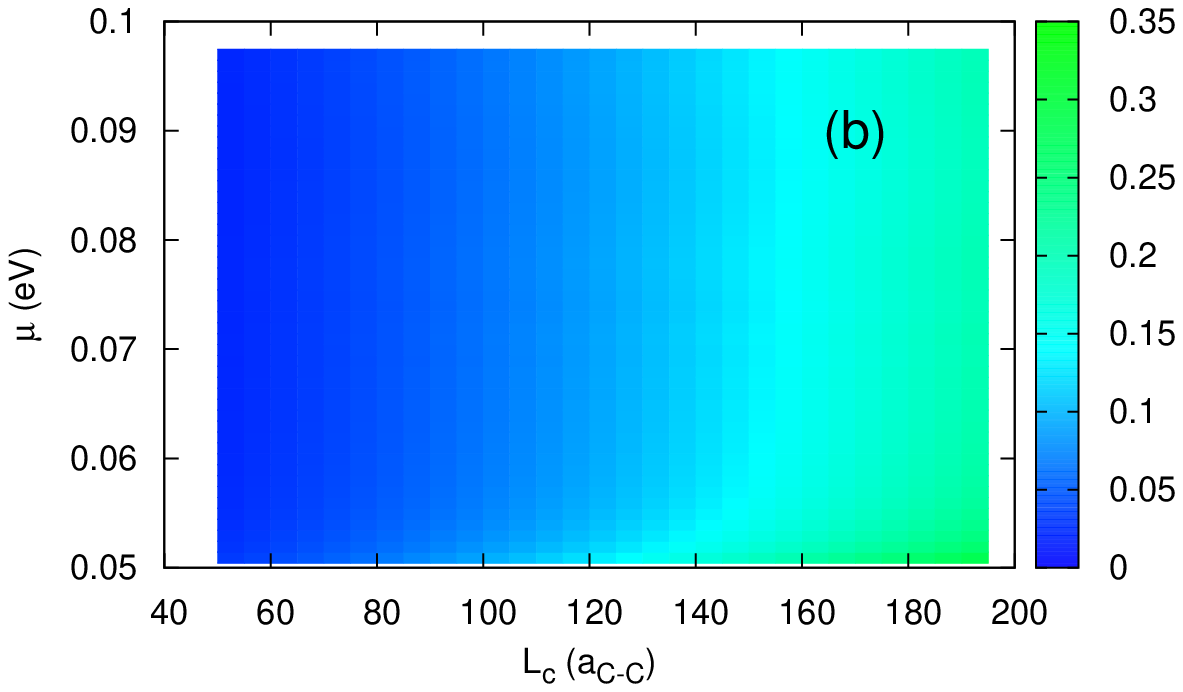}
  \caption{
    (color online) (a) Spin-flip relative conductance $X_s$, calculated at $T=1.8$~K, 
    for a system with \emph{direct} bias of $\Delta=0.1$~eV, and \emph{inverse} bias (b). 
    $\mu$ refers to the electrochemical potential with respect to midgap of the bilayer bands, 
    as shown in Fig.~\ref{fig:1}(c).   
  }
  \label{fig:cond}
\end{figure}

\section{Conductance}
A readily measurable property of the system is its conductance.
We have therefore calculated the 2D conductance of the bilayer system with FO in the C region.
In particular we are interested in the spin-flipped relative conductance $X_s=G_{\uparrow\downarrow}/G$, 
with $G=G_{\uparrow\downarrow}+G_{\uparrow\uparrow}$.
This is a measure of the efficiency of spin control in the proposed device.
The 2D  two-terminal conductance, which is defined by $dI_x = G_x dV_{DS}$, is expressed as
\begin{equation}
  G_x = \frac{g q^2}{(2\pi)^2} \int_{BZ} dk_x dk_y T_{k_x,k_y} v_x \frac{df(E_{k_x,k_y}-\mu)}{dE},
\end{equation}
with $g=4$ (accounting for the valley and spin degeneracy), $v_x$ the group velocity along the transport direction, 
$f(E)$ the Fermi-Dirac distribution function and $\mu$ the electrochemical potential. 
Integration is performed over the Brillouin zone.

In Fig.~\ref{fig:cond} we show the spin-flip relative conductance $X_s$ as a function of the electrochemical 
potential for $L_C$ from $50 a_{C-C}$ to $200 a_{C-C}$, at the temperature of $1.8$~K.
Fig.~\ref{fig:cond}(a) is computed with a \emph{direct} gate bias of 
$\Delta=0.1$~eV, while (b) shows the case with the \emph{reverse} bias.
The resonance present in (a), corresponding to an electrochemical potential for which 
the spin-flip affects more than  $80\%$ of transmitted electrons, 
is completely absent in (b), where electrons tend to preserve their original spin.
This calculation clearly demonstrates that we are able 
to control the spin-flip of carriers traveling through the system by changing the gate bias.

The total conductance is thermally activated as $\mu$ approaches the bottom of the conduction band of the LS region.
As $\mu$ enters the 3-states spectral region in Fig.~\ref{fig:1}(c) the conductance spin properties 
are dominated by the behavior of the transmission probability in Fig.~\ref{fig:tunn}(a).
This leads, with a \emph{direct} bias, to a pronounced resonance of $X_s$.
A fundamental factor for the appearance of this resonance is that the spin differential transmission resonance 
in Fig.~\ref{fig:tunn}(a) is almost isotropic for small $k_y$, similar to the LS dispersion curve, 
deviating only for large $k_y$.
Therefore, it is possible, with the appropriate electrochemical potential, to adjust the Fermi level 
to this transmission resonance. 
The states relevant to the conductance will therefore be quite well collimated on the spin-flip transmission 
resonance.
As expected, an increase of the temperature leads to a broader state population, gradually blurring away this feature.
\begin{figure}[t]
  \centering
  \includegraphics[width=7cm]{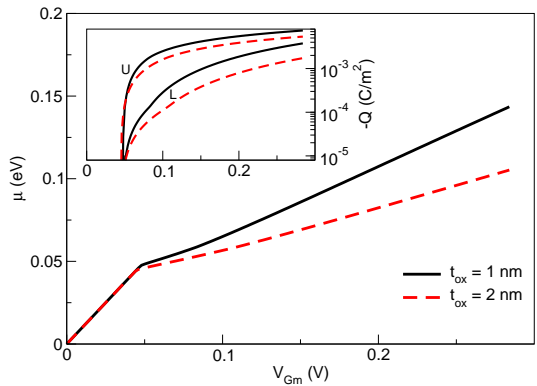}
  \caption{
    (color online) Electrochemical potential as a function of $V_{Gm}$ for a double gate graphene bilayer 
    in quasi-equilibrium condition, at a temperature $T=1.8$~K, 
    calculated for $t_{ox}=1$, $\Delta V_G=0.023$~V and $t_{ox}=2$~nm, $\Delta V_G=0.034$~V.
    In the inset the charge accumulated on the U and L graphene layers in the two cases is shown.    
  }
  \label{fig:electr}
\end{figure}

\section{Self-consistent analysis}
To provide an indication of the real control that the gates exert on the system, 
and therefore of the observability of the phenomenon, we performed a self-consistent electrostatic analysis.
Indeed, we can fix the absolute value of the chemical potential, 
but we cannot set the difference between the electrochemical potential and the bilayer graphene midgap.
In other words, the value of the potential of the U and L layers of the graphene bilayer is the result 
of the self-consistent calculation, which depends on the gate voltages, taking into account the capacitive 
coupling with the gates.
We study a double gate FET in which we can independently fix the top and back gate voltages ($V_{GT}$ and $V_{GB}$).
Alternatively we can give the average gate potential $V_{Gm}=\frac{V_{GT}+V_{GB}}{2}$, 
which is responsible for rigidly shifting the bands (and therefore varying the electrochemical potential $\mu$
with respect to midgap), and the gate voltage difference $\Delta V_G = V_{GB}-V_{GT}$ which opens up the 
semiconducting gap of the graphene bilayer.
To describe the electrostatics of the system we apply to the graphene bilayer the plane 
capacitor model described in Refs.~\onlinecite{castro2008,cheli2009}.
Another way to describe the charge on the U and L plane is band filling.
In fact, the occuppation of each one of the graphene bilayer state is described by the 
Fermi-Dirac distribution, and the charge it carries can be distributed on the U 
and L plane according to $P_U$ and $P_L$.
The two descriptions of the system, electrostatics and statistics, 
should be consistent and their simultaneous solution fixes the U and L potentials, 
and therefore $\mu$.

We focus our calculation on a system with $\Delta \approx 0.1$~eV, and analyze the control on 
the electrochemical potential with respect to the midgap of the graphene bilayer.
In Fig.~\ref{fig:electr}, we show $\mu$ as a function of the average gate potential $V_{Gm}$, 
where a potential difference $V_{GB}-V_{GT}=0.023$~V and $0.034$~V has been applied for 
$t_{ox}=1$ and $2$~nm respectively (values which lead to $V_L-V_U \approx 0.1$~V).
When the device is empty the electrochemical potential linearly increases with $V_{Gm}$.
As the electrochemical potential reaches the bottom of the conduction band, 
we can observe an abrupt change of slope.
As the charge accumulates in the device, the variation of the electrochemical potential 
becomes more difficult, due to the increase of the quantum capacitance of the system~\cite{luryi1988}.
The spin-flip resonance region is easily reached with the tight double gate structure adopted here, 
which optimizes the electrostatic control.
The considered oxide thicknesses are obtained with state-of-the-art semiconductor technology,
and high-dielectric-constant oxides (the so called high-K dielectrics) can allow even better
electrostatic control.
In the inset of Fig.~\ref{fig:electr} we show the charge accumulated on the U and L plane.
The charge shows an activation behavior in correspondence with the value of $V_{Gm}$ for which the 
electrochemical potential reaches the conduction band.

\section{Conclusion}
We have demonstrated that bilayer graphene FETs, in which a ferromagnetic 
insulator is used as a gate dielectric, is an interesting system for spin manipulation.
In particular, we have shown that a good electric control of spin rotation can be 
achieved even in a 2D system without lateral confinement, at low temperature.
We show that by switching between a \emph{direct} and a \emph{reverse} gate polarization, 
we can modulate the ratio of spin-flipped transmitted carriers from more than $80\%$ to less than $20\%$.
Therefore, the system itself acts as a tunable spin-flipping device and offers the possibility to devise
spin-FETs based on bilayer graphene, 
exploiting the exchange proximity interaction with a ferromagnetic insulator, 
instead of the rather weak intrinsic spin-orbit coupling.

\begin{acknowledgements}
P. M. and P. R. acknowledge financial support from the DFG via the Emmy Noether program.
G. I. acknowledges financial support from the EC through the FP7 Nanosil NoE (contract n.~216171). 
\end{acknowledgements}

%%%%%%%%%%%%%%%%%%%%%%%%%%%%%%%
\bibliography{bibf}

\end{document}